# Growth monitoring with sub-monolayer sensitivity via real time thermal conductance measurements.

P. Ferrando-Villalba[1], D. Takegami[1], Ll. Abad[2], J. Ràfols-Ribé[1], A. Lopeandía[1*], G. Garcia[1], J. Rodríguez-Viejo[1]

[1] Grup de Nanomaterials i Microsistemes, Dep. Física, Universitat Autònoma de Barcelona, 08193 Bellaterra, Spain.

[2] Institut de Microelectrònica de Barcelona- Centro Nacional de Microelectrònica – CSIC, Cerdanyola del Vallès, 08193, Spain

## Abstract

Growth monitoring during the early stages of film formation is of prime importance to understand the growth process, the microstructure and thus the overall layer properties. In this work, we demonstrate that phonons can be used as sensitive probes to monitor real time evolution of film microstructure during growth, from incipient clustering to continuous film formation. For that purpose, a silicon nitride membrane-based sensor has been fabricated to measure in-plane thermal conductivity of thin film samples. Operating with the $3\omega$-Völklein method at low frequencies, the sensor shows an exceptional resolution down to $\Delta(\kappa \cdot t) = 0.065 \frac{W}{m \cdot K} \cdot nm$, enabling accurate measurements. Validation of the sensor performance is done with organic and metallic thin films. In both cases, at early stages of growth, we observe an initial reduction of the effective thermal conductance of the supporting amorphous membrane, $K$, related with the surface phonon scattering enhanced by the incipient nanoclusters formation. As clusters develop, $K$ reaches a minimum at the percolation threshold. Subsequent island percolation produces a sharp increase of the conductance and once the surface coverage is completed $K$ increases linearly with thickness The thermal conductivity of the deposited films is obtained from the variation of $K$ with thickness.

email: aitor.lopeandia@uab.cat

## Introduction

Monitoring the first stages of thin film growth is of key importance to understand and thus tune the properties of grown layers. Critical microstructure features such as grain size, morphology, crystal orientation, nature of grain boundaries and surface morphology, are defined during the early growth process. Real-time measurements have proven their potentiality to understand the growth dynamics, either for thin films or nanoparticles deposited on surfaces. Actually, in-situ diagnostics during growth with monolayer sensitivity have already been performed by a variety of techniques such as wafer curvature measurements mapping the stress evolution[1], ellipsometry[2], X-ray reflectivity[3,4] and resistance-based measurements. Low or medium energy electron diffraction (LEED, MEED) are also reliable tools to monitor 2D ordering during epitaxial growth[5,6]. Among all, electrical measurements are very powerful since i) the electrical resistance in metallic thin films[7] may vary orders of magnitude above the percolation threshold , and ii) drastic electrical conductivity changes during the initial growth stages can identify phase transformations, such as amorphous-to-crystal transition in Mo films[8]. Unfortunately, although simple and accessible, this approach is limited to metallic or highly conductive layers, precluding the analysis of organic or insulating materials. In contrast, phonons are a more generic probe, extremely sensitive to film structure thanks to their larger mean free path compared to electrons. However, real-time thermal conductance measurements during film growth are much scarcer, mainly due to the technical challenges associated.

Additionally, the potential application of nanomaterials, thin films and nanostructures in heat management and efficient thermoelectric devices has boosted the necessity to perform accurate the thermal conductivity measurements at the nanoscale. In particular, phonon engineering in low-dimensional materials has appeared as the most effective approach to enhance the thermoelectric figure of merit[9–11] through the reduction of thermal conductivity. The implementation of novel nanomaterials designs has encouraged the development of new sensors and methodologies, enabling accurate determination of thermal conductivity in low dimensional architectures. Whether based on optical[12] or electrical[13] signals, these novel thermal sensors and associated methodologies have allowed in-plane[14,15] and out-of-plane[16,17] thermal conductivity measurements, of nanowires and thin films, with outstanding nanometre spatial resolution[18,19]. A remarkable contribution to the field was achieved by Völklein et al. in 1990 who developed a suspended membrane-based sensor using a long and thin Pt electrical transductor operated in DC to measure in-plane thermal conductivity of thin films[15]. More recently, Sikora et al. went a step further in improving this technology by combining the Völklein geometry with the AC 3$\omega$-method, reaching exceptional thermal conductance sensitivity, $\frac{\Delta K}{K} \cong 10^{-3}$[20,21].

In parallel, ex-situ thermal probe studies performed on thin films[22,23] have shown that the thermal conductivity of a grown material is conditioned by the thermal conductance loss of the substrate induced both by interfacial scattering of phonons in the in-plane measurements[22] and by thermal boundary resistance in the out-of-plane measurements[23]. Accordingly, the thermal conductivity of the thin layer cannot be simply calculated via the differential measurement of the thermal conductance of the whole sample (film + membrane/substrate) and a reference (only membrane/substrate), but must be calculated using a set of thermal conductance measurements performed at different film thicknesses. Up to date, most of these measurements have only been performed ex-situ, evaluating the temperature dependence of the thermal conductivity $k$ for each singular thickness. To our knowledge, real-time studies during growth at early stages taking into account the impact of the microstructure on phonon scattering have not been previously reported. It is worth noting that although Völklein and Starz[24] already demonstrated in 1997, that a Völklein-like sensor, operating in DC mode, could perform in-situ measurement of thermal conductance in thin films, those measurements were limited to metallic films thicker than 1 µm, on the contrary to the sensor presented here.

In this paper we present i) an improved 3$\omega$-Völklein technique allowing highly sensitive conductance measurements ($\Delta K/K < 10^{-3}$), ii) the sensor optimization through FEM (Finite Element Modelling) thermal analysis, iii) a novel study of the thermal evolution of the sensor during growth and iv) real-time measurements of the thermal conductance during growth of both highly insulating and conductive (electrically and thermally) materials with sub-monolayer sensitivity. Proof-of-concept is achieved by analysing two different materials: a metal, such as In (indium), and an organic conductor, N,N′-Bis(3-methylphenyl)-N,N′-diphenylbenzidine (TPD), often used as hole injector in OLEDs. Complementary characterization techniques (SEM, AFM and electrical measurements) are used to correlate conductance features with sample morphology throughout the growth.

## Experimental Section

### Sensor Design and Simulation

The sensor developed here consists in a long and thin Pt line deposited on a suspended SiN$_x$ membrane and connected in a 4-wire configuration (Fig. 1a,b), along with an external Pt line. The thermal conductance of the whole membrane is determined with the centred Pt line (normal operation), while both lines, centred and external, are needed to measure exclusively the thermal conductance of the membrane volume portion between them. Details of the sensor microfabrication can be found in the Supplementary Information (SI), in Fig. S1.

In normal operation, the thermal conductance from the central Pt line to the substrate is calculated by using the 1D Fourier law and assuming that the external line is much more conductive than the SiN$_x$ beneath, which yields equation 1

$$K = k_{SiN_x} t_{SiN_x} \left(\frac{L}{l} + \frac{L}{l-w}\right) \qquad (1)$$

where $t_{SiN_x}$ and $k_{SiN_x}$ are the thickness and the thermal conductivity of the silicon nitride respectively, $L$ is the length of the Pt strip between the voltage probes, $w = 4.5$ µm is the width of the Pt strip and $l = 0.002$m is the distance between the Pt strip and the substrate.

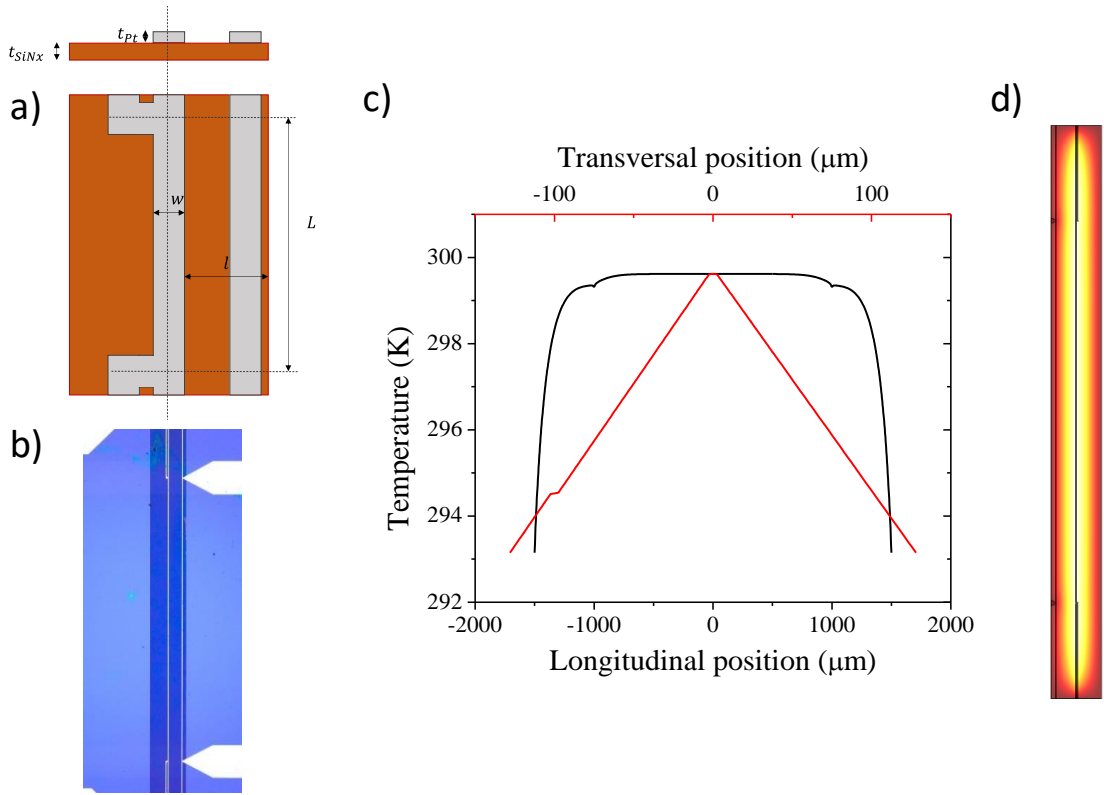

**Fig. 1**a) Scheme of the sensor (side and top views), consisting on a suspended membrane with a Pt strip on it. b) Optical micrograph of the sensor. c) Longitudinal and transversal temperature distributions of the sensor in a FEM DC study. d) Surface temperature distribution of the sensor (brighter is hotter) in a FEM DC study.

We have modelled several sensor geometries using Finite Element Model (FEM) with COMSOL software in order to ensure temperature homogeneity between the voltage probes of the central strip. In the first place, we have introduced voltage probes to measure the temperature oscillations of the central part of the strip, which has a flatter temperature profile than the borders. This simulation reproduces a steady state by feeding the Pt strip with a DC current. The optimum sensor (considering the resolution limits of the available photolithographic system and the structural stability of the $SiN_x$ membrane) consists in a long and narrow $SiN_x$ membrane (3mm x 250 μm x 180 nm) that supports the two Pt sensing strips (3mm x 5 μm x 100 nm) with voltage probes separated by 2 mm. A key point in this design is the close proximity of the voltage probes and the central Pt strip, minimizing heat leakage and any temperature drop along the strip. The simulated profile of the longitudinal temperature (Fig. 1c) shows a large and flat central plateau where the inhomogeneity is lower than 3% of the total temperature rise. We notice two tiny temperature depressions at ±1000 μm, coinciding with the voltage probe locations. The transversal profile exhibits a constant slope only perturbed by minimal flat segments corresponding with the position of the Pt strips. With this sensor, the thermal conductivity of the SiN$_x$ layer can be measured with an accuracy of 0.1% in the absence of other uncertainty sources. Main parameters used and more information about the modelling can be found in the SI (Fig. S2, S3).

**Thermal analysis during film deposition**

Based on Sikora et al. developments[20,21], we deduced a mathematical expression to extract the thermal conductivity from the measured thermal conductance at different thicknesses. When the central Pt line is fed with an oscillating current $I = I_0 \sin \omega t$, the amplitude of the temperature oscillations induced can be calculated with the 1D heat equation (assuming that the sensor is infinitely long), which after some approximations (See SI) yields:

$$\Delta T_{2\omega} = \frac{P_0}{K\sqrt{1+\omega^2\left(4\tau^2+\frac{2l^4}{3D^2}+\frac{4\tau l^2}{3D}\right)}} \quad (2)$$

where $P_0 = \frac{I_0^2 R}{2}$, $D = \frac{k_{SiN_x}}{\rho_{SiN_x} \cdot c_{SiN_x}}$, $\tau = \frac{C_p}{K}$ and $C_p = wL(\rho_{SiN_x} \cdot c_{SiN_x} \cdot t_{SiN_x} + \rho_{Pt} \cdot c_{Pt} \cdot t_{Pt})$. Here $\rho_{SiN_x,Pt}$ is the density of the SiN$_x$ membrane and the Pt strip, $c_{SiN_x,Pt}$ is their specific heat capacity, $C_p$ is the heat capacity of the strip and the portion of SiN$_x$ beneath it, $\tau$ is the thermal time constant of the sensor, $D$ is the thermal diffusivity of SiN$_x$, $R$ is the Pt strip electrical resistance between the voltage probes and $I_0$ is the current amplitude. From equation (2), the apparent thermal conductance can be calculated as $K_{2\omega} = P_0/\Delta T_{2\omega}$, which resembles $K$ at low frequencies.

When a thin-film sample grows on the SiN$_x$ membrane, the parameters in equation (2) may vary as $k, c$ and $\rho$ will no longer correspond solely to the SiN$_x$ membrane but to the combination of the deposited film (from now on called "sample") and the substrate membrane. In this case, an effective value for these magnitudes can be calculated (assuming a vertical growth of the film) by pondering the different values as a function of the film thickness, $t_{smp}$:

$$k_{eff} = \frac{k_{SiN_x} \cdot t_{SiN_x} + k_{smp} \cdot t_{smp}}{t_{SiN_x} + t_{smp}} \quad (4)$$

$$c_{eff} = \frac{c_{SiN_x} \cdot t_{SiN_x} + c_{smp} \cdot t_{smp}}{t_{SiN_x} + t_{smp}} \quad (5)$$

$$\rho_{eff} = \frac{\rho_{SiN_x} \cdot t_{SiN_x} + \rho_{smp} \cdot t_{smp}}{t_{SiN_x} + t_{smp}} \quad (6)$$

$$D_{eff} = \frac{k_{eff}}{\rho_{eff} \cdot c_{eff}} \qquad (7)$$

Here, $k_{smp}$, $c_{smp}$ and $\rho_{smp}$ are the sample thermal conductivity, specific heat capacity and density. In the same way, the extrinsic values $K$ and $C_p$ will vary, as well as the thermal time constant $\tau$, as:

$$K(t_{smp}) = (k_{SiN_x} \cdot t_{SiN_x} + k_{smp} \cdot t_{smp})\left(\frac{L}{l} + \frac{L}{l-w}\right) \qquad (8)$$

$$C_p(t_{smp}) = wL(\rho_{SiN_x} \cdot c_{SiN_x} \cdot t_{SiN_x} + \rho_{Pt} \cdot c_{Pt} \cdot t_{Pt} + \rho_{smp} \cdot c_{smp} \cdot t_{smp}) \qquad (9)$$

$$\tau' = \frac{C_p(t_{smp})}{K(t_{smp})} \qquad (10)$$

If the measuring frequency is low ($\omega \ll \frac{3D_{eff}}{8l^2}$), then the measured $K_{2\omega}$ resembles $K$ and the derivative of equation (8) can be used to calculate the thermal conductivity of the sample film by measuring in real-time the thermal conductance during growth ($K_{2\omega}(t_{smp})$), as shown in equation (11).

$$\frac{dK}{dt_{smp}} = k_{smp}\left(\frac{L}{l} + \frac{L}{l-w}\right) \rightarrow k_{smp} = \frac{\frac{dK}{dt_{smp}}}{\left(\frac{L}{l} + \frac{L}{l-w}\right)} \qquad (11)$$

However, if a higher current angular frequency $\omega$ is used, $k_{smp}$ can be extracted by fitting the measured $K_{2\omega}$ with equation (12) using the thickness-dependent parameters (equations (7) - (10)):

$$K_{2\omega}(t_{smp}) = K(t_{smp})\sqrt{1 + \omega^2\left(4\tau'^2 + \frac{2l^4}{3D_{eff}^2} + \frac{4\tau' l^2}{3D_{eff}}\right)} \qquad (12)$$

In Fig. 2a the calculated $K_{2\omega}$ is plotted for current frequencies of 0 Hz (DC), 1 Hz and 3 Hz (the parameters used are listed in the figure caption). If the conductance is monitored with a current at 1 Hz, the apparent thermal conductance $K_{2\omega}$ is very similar to $K$ throughout the deposition. However, at 3 Hz, there is an evident difference in the slope of the curves: Although the absolute value of the apparent thermal conductance only varies from 97% to 93% of $K$ (Fig.2b), the slope of $K_{2\omega}$ is up to 40% higher than the one of $K$ (Fig.2a). Generally, measuring with higher frequencies increases the dependence of $K_{2\omega}$ with intrinsic properties of the sample that may not be well known (like $c_{smp}$ or $\rho_{smp}$), hindering the determination of $k_{smp}$.

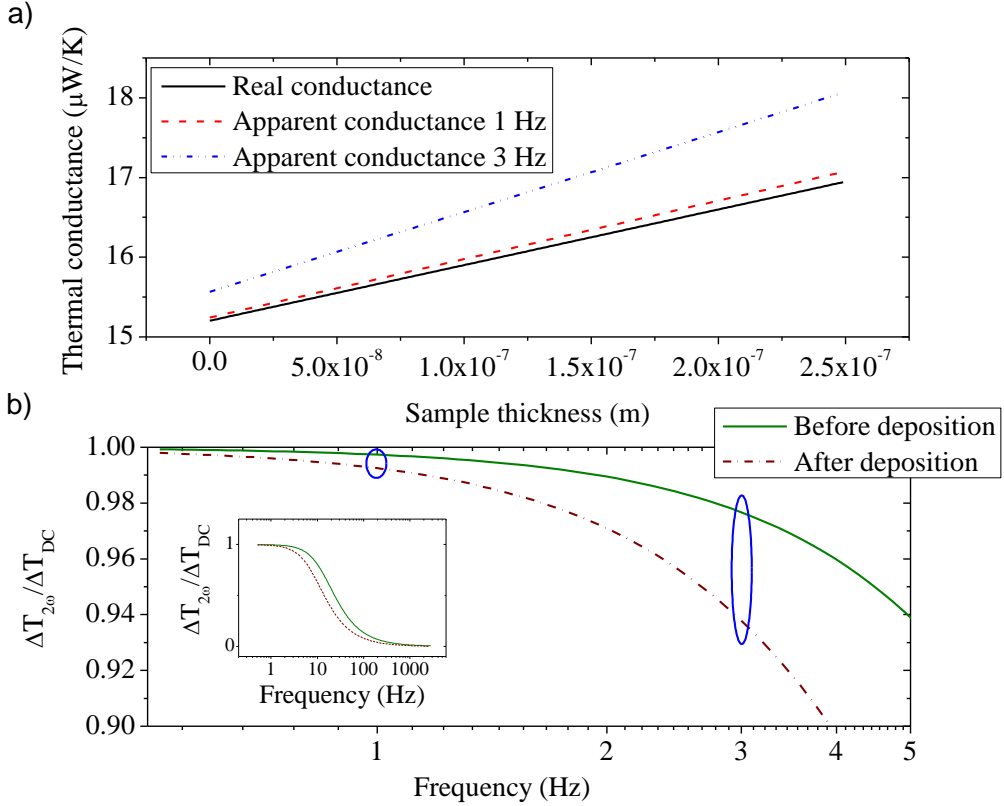

**Fig.2** a) Calculated real thermal conductance (black) and apparent thermal conductance modelled at 1 Hz (red) and 3 Hz (blue) during the deposition of a layer. b) Calculated frequency dependence of the temperature oscillations in the sensor before and after depositing a 250 nm film (the inset is a zoom out). The material properties used in both plots are the following: $k_{SiN_x} = 2.65 \text{Wm}^{-1}\text{K}^{-1}$, $c_{SiN_x} = 0.7 \text{ J·K}^{-1}\text{·Kg}^{-1}$, $\rho_{SiN_x} = 3.18 \text{g/cm}^3$, $k_{Pt} = 33 \text{Wm}^{-1}\text{K}^{-1}$, $c_{Pt} = 0.133 \text{J·K}^{-1}\text{·Kg}^{-1}$, $\rho_{Pt} = 21.45 \text{ g/cm}^3$, $k_{smp} = 0.21 \text{Wm}^{-1}\text{K}^{-1}$, $c_{smp} = 1.05 \text{J·K}^{-1}\text{·Kg}^{-1}$ and $\rho_{smp} = 1.08 \text{ g/cm}^3$. Structural parameters are $t_{SiN_x} = 180$ nm, $t_{Pt} = 110$ nm, $l = 123 \mu m$, $w = 5 \mu m$ and $L = 2000$ μm.

## Experimental setup

The sensor is introduced in a high vacuum chamber equipped with an effusion cell that enables a good control of the evaporation rate. A previously calibrated quartz crystal microbalance is located nearby the sensor to monitor the layer growth rate with a precision of 0.01 Å/s. The temperature of the sample was controlled with a custom-made PID system that reads the temperature of a Pt100 and provides heat through a Kapton heater, yielding a temperature control with fluctuations smaller than 0.003K, from 77 K up to 400 K.

The experiment is performed by feeding two sensors (sample and reference) with a current wave of a given amplitude and frequency, generating a voltage drop in each sensor. The voltage signals from both sensors, as well as the differential voltage between them, were subtracted using the low-noise amplifiers with gains $G_{smp}$, $G_{ref}$ and $G_{diff}$, respectively. The reference sensor is a twin (equal to the sample one) non-suspended sensor which produces no self-heating, and is used to subtract the $1\omega$ component of the sample sensor. Thus, the $3\omega$ component of the differential voltage is only produced in the sample sensor, and owing to the cancellation of the $1\omega$ voltage, it could be amplified with a gain $G_{diff} = 75$. The main benefit of using a twin sensor (instead of a variable resistance) as a reference for the differential measurement is that, if the temperature of the sample holder varies, the resistance of both sensors will change hand-in-hand, making unnecessary to build a control system for the cancellation of the $1\omega$ voltage. The exact electronics used in the measurement of the different voltage signals are detailed in the SI (Fig. S3).

From the measured voltage signals, the resistance $R_{smp}$ and the temperature oscillations $\Delta T_{2\omega}$ can be calculated as:

$$R_{smp} = \frac{V_{1\omega}}{I_0 \cdot G_{smp}} \qquad (13)$$

$$\Delta T_{2\omega} = \frac{2V_{3\omega}}{I_0 \frac{dR_{smp}}{dT} \cdot G_{diff} \cdot G_{smp}} \qquad (14)$$

where $V_{1\omega}$ and $V_{3\omega}$ are the $1\omega$ and $3\omega$ voltage components measured in the sample sensor, and $\frac{dR_{smp}}{dT}$ is the slope of the sample resistance as a function of the temperature.

## Results

**Sensor performance test**

Initially, the self-heating of the Pt sensor is determined by measuring both the $3\omega$ voltage and the variation of the $1\omega$ voltage with $\omega = 2\pi\ rad/s$ (Fig.3a). Since the frequency is very low, both signals yield identical self-heating, but as can be seen in Fig. 3a, the self-heating calculated with the $3\omega$ voltage ($\Delta T_{2\omega}$) is less noisy than the one calculated with the $1\omega$ voltage ($\Delta T_{DC}$). Also, the slope of $\log \Delta T_{2\omega}$ vs $\log I_0$ has a value very close to 2, demonstrating that the self-heating depends on the square of the current.

$\Delta T_{2\omega}$ has been measured in a wide frequency range (1 Hz to 2000 Hz) and compared with finite element simulations in a time-dependent study (See SI), as can be seen in Fig. 3b. The high coincidence between both datasets suggests that the behaviour of the sensor is purely driven by heat transport physics. This is an important difference from the device presented by Sikora et al.[20], where the use of a NbN strip sensor allowed measurements at very low temperatures, but produced non-negligible electrical effects due to the high electrical impedance of that material.

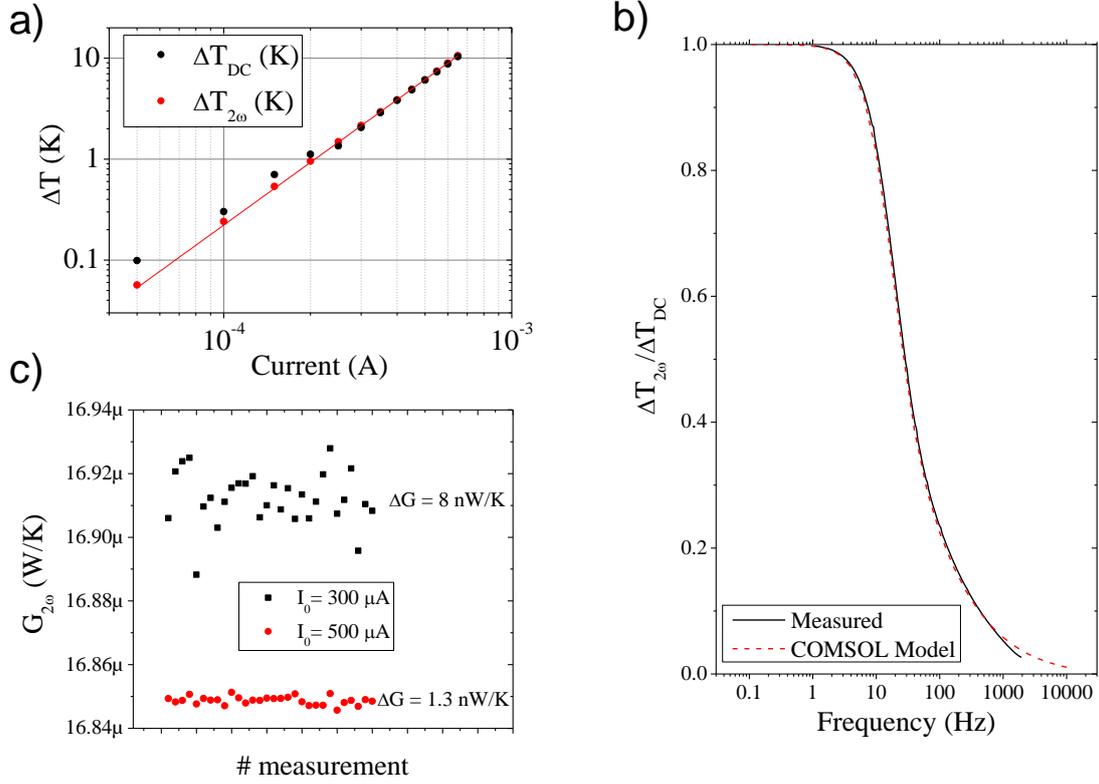

**Fig.3** a) Self-heating of the sensor measured with the $3\omega$ and $1\omega$ voltages ($\omega = 1$Hz). b) Temperature oscillations as a function of the frequency. The measured values are compared with the simulated ones using the COMSOL model. c) Measurement of the uncertainty in the thermal conductance $\Delta K$ refers to the standard deviation of the data. Red circles and black squares correspond to data obtained with input currents of 500 and 300 µA respectively, with $\omega = 1$Hz.

Finally, as shown in Fig.3c, the uncertainty of the conductance measurement has been determined at two different current intensities, 300 µA and 500 µA, which generated temperature amplitudes of about 2 K and 6 K and produced a standard deviation of data equal to 8 nW/K and 1.3 nW/K respectively. This accuracy in the thermal conductance provides an extremely high resolution in the product of the thermal conductivity and the thickness, $\Delta(k \cdot t) = 0.065 \frac{W}{m \cdot K} \cdot nm$. Although there is a decrease in the data dispersion for 500µA, there is also a noticeable reduction in the average value of the thermal conductance, from 16.91 µW/K to 16.85 µW/K that may be justified by a small reduction of the sensor dR/dT due to the higher average heater temperature.

The thermal conductivity of the SiN$_x$ membrane was evaluated for temperatures between 80 K and 230 K, as shown in Fig. 4 (black squares). The results are very similar to the values found in the out-of-plane direction using the 3$\omega$ method on a similar SiN$_x$ membrane (red circles in Fig. 4). The latter values were obtained through a differential measurement of two samples with thicknesses of 180 nm and 450 nm, granting that the thermal boundary resistances between the film and the substrate are cancelled. Thus, the similarity of both the in-plane and the out-of-plane values confirm that there are not substantial phonon size effects, or anisotropy, in our layers. In both cases, the variation of $k$ with temperature show also a similar tendency than data presented by Sikora et al[20,21] (continuous line in Figure 4). The discrepancy in the absolute values may account for density or stoichiometry variations related to the different growth film characteristics in each work. Huge differences in $k$ variation with temperature are observed when comparing our data with Sultan et al.[25] work (blue triangles in Fig.4). We suggest, as will be discussed later, that this difference is probably related to fact that they use a nanocrystalline membrane.

The measurements performed in this work have an extremely low variability owing to the high sensitivity of the method. The main source of uncertainty in the in-plane measurement is the precise determination of SiN$_x$ membrane thickness, which was measured throughout the wafer with a contactless optical profilometer, yielding a standard deviation of 1% of the 180 nm averaged thickness. However, wet etchings at the final steps of microfabrication process could slightly reduce this thickness up to 10 nm, inevitably increasing the uncertainty up to 5%.

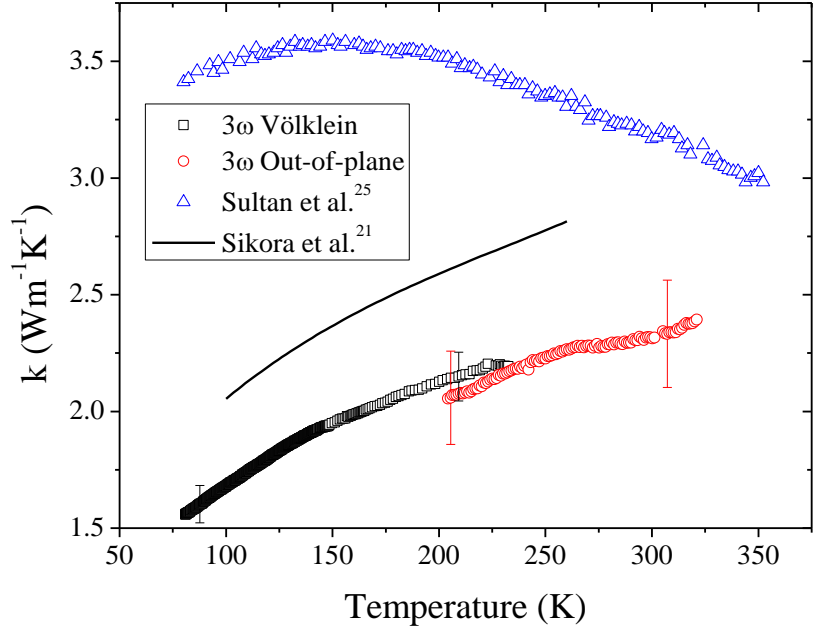

**Fig. 4** Thermal conductivity values measured at different temperatures with the 3ω-Völklein Method (in-plane direction) and the 3ω method (out-of-plane direction). Data from Sultan et al.[25] and Sikora et al.[21] is also included for comparison.

**Real-time measurement of thin-film growth: sensor proof of concept**

Organic thin-film layers

We initially measure the thermal conductance of the membrane as a function of the TPD thickness deposited at 266±2K, as shown in Fig.5a. The temperature uncertainty comes from the thermal oscillations produced by the current wave of amplitude 300 μA. The evaporation was carried out below $10^{-7}$ mbar, by heating up an effusion cell up to 200°C, which yielded a stable deposition rate of 0.29nm/s. From the slope of the curve $K_{2\omega}(t_{smp})$ and applying equation (9), we calculate the thermal conductivity of the TPD layer, resulting in $k_{smp} = (0.153 \pm 0.001)$Wm$^{-1}$K$^{-1}$. However, if the thermal conductivity is calculated by fitting equation (12) we obtain $k_{smp} = (0.145 \pm 0.001)$ Wm$^{-1}$K$^{-1}$. As discussed previously both results are comparable due to the low frequency used in the measurement. The low value of the thermal conductivity is an indication of the glassy character of the TPD layers, as confirmed by calorimetric measurements showing a clear signature of the glass transition temperature (SI, Fig.S6).

As indicates Fig. 5a, at the early stages of the deposition, i.e. very low thicknesses, we observe a decrease in the thermal conductance of around 1.2 % of its initial value, while afterwards $K_{2\omega}$ roughly increases linearly with thickness. In this particular measurement, film growth was stopped at 340 nm, yielding a constant value of the thermal conductance after this particular point.

Fig.5b,c show in more detail the initial stages of the evolution of $K_{2\omega}$ versus thickness for two TPD films deposited at 267 K and 304 K, both with a low growth rate of 0.02 nm/s. In both cases, we identify four different regions where thermal conductance follows different trends with thickness. In region I, the overall conductance decreases abruptly following an exponential behaviour, while it decreases slowly in Region II until a minimum point is reached. Region III covers the thickness range where $K_{2\omega}$ increases slowly up to a point where a linear regime of the conductance with thickness is attained and labelled as Region IV. The extent of each of these regions is represented in Fig. 5b,c by dashed red lines, and as is clearly seen, vary from one sample to another, confirming the importance of deposition temperature as will be discussed later.

To understand and interpret the variation of $K_{2\omega}$ with thickness, we also performed ex-situ analysis of the film morphology either by SEM or by AFM described in Fig. S4a of the SI. Complementary electrical characterizations were also performed, as shown in Fig. 5b inset, to corroborate the hypothesis described later on. Surface morphology evolution with thickness showed that TPD do not wet the $SiN_x$ membrane surface, as it grows in 3D islands in the early growth stages. For films deposited at 267K and according to the conductance evolution, island coalescence seems to happen at a nominal film thickness of 2.6 nm while complete surface coverage seems to take place at 14 nm. Ex-situ analysis of growth regimes is precluded by the evidence of dewetting of TPD at temperatures well below the glass transition temperature[26]. The fast film dynamics at temperatures below the glass transition temperature makes in-situ analysis an indispensable tool to gain a better understanding of film formation in molecular glasses.

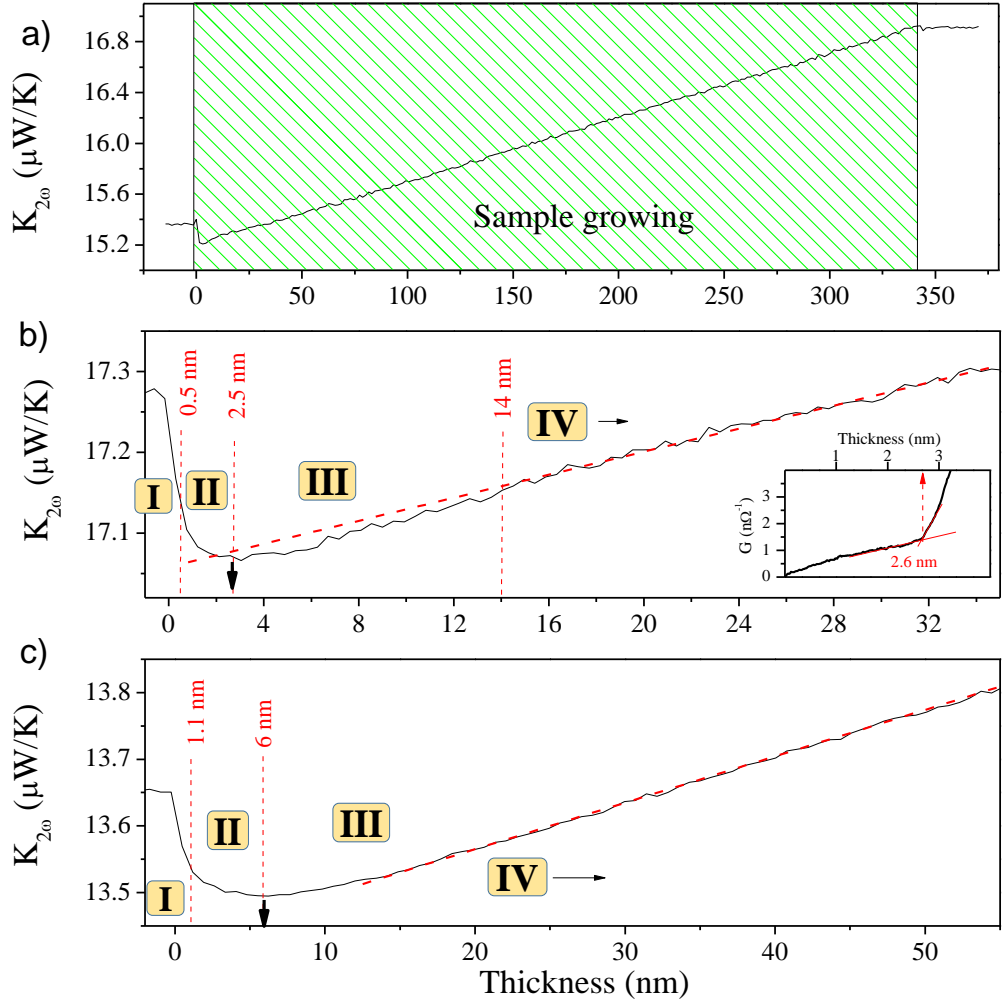

**Fig. 5** Thermal conductance vs thickness during TPD growth: a) $T_{dep}$=267±2 K. Growth rate = 0.29 nm/s. b) $T_{dep}$=267±2 K. Growth rate = 0.02 nm/s. c) $T_{dep}$ = 304±2 K. Growth rate = 0.02 nm/s. In graphs b) and c) data is box averaged with 10 points/box. The main regions of film growth are separated by dashed red lines. Region I located between 0 nm and the first vertical line correspond to nucleation and isolated island formation; region II to island coalescence; region III to percolation across the layer and region IV to vertical growth of a continuous layer. The slight difference in the initial conductance between all the graphs is mainly due to the use of different sensors/membranes for the experiments. The black downward arrow marks the percolation threshold (separation between regions II and III). The inset of b) shows the abrupt variation of the electrical conductance at the percolation threshold (nominal thickness of 2.6 nm) of the TPD film. This value coincides with the minimum of the thermal conductance (black arrow) in graph b).

Before relating the large drop observed in region I with film microstructure, we performed few tests to rule out any potential experimental artefact that could have led to temperature variations of the sensor. As extra heat originated by condensation of molecules on the substrate is not modulated at angular frequency $\omega$ or any of its harmonics and therefore, does not affect the $3\omega$ component of the voltage, the decrease of conductance cannot be attributed to a sensor temperature variation due to this phenomenon. Also, we ruled out any effect of the sample emissivity variations during TPD growth on the membrane, since a measurement at 81 K, where radiation effects should be substantially lower, showed a similar drop in thermal conductance (Fig. S5 in SI). The reduction of thermal conductance observed in the early stages of growth seems thus to be due to phonon-related phenomena, although covering the surface with TPD is not expected to affect the thermal conductance of the $SiN_x$ membrane since it is already a disordered structure with an average phonon mean free path of the order of the interatomic distance.

Nevertheless, a previous work by Sultan et al.[22,25] already showed similar behaviour ($K$ decrease at early growth stages with a subsequent increase for larger thicknesses) by performing ex-situ measurements of thermal conductance versus thickness of metal and alumina deposited on $SiN_x$. Nonetheless, our data differs from this previous work in several important aspects: i) the magnitude of the conductance drop is much lower and ii), our measurements, are carried out in-situ with a much higher conductance sensitivity, providing a much clear signature of the various growth regimes.

Our thermal conductance initial drop, although much higher than the uncertainty of the measurement, only amounts 1.2%, compared to 5-10% in Sultan's work. These authors estimated that 40%-50% of the total $K$ was due to long-wavelength with long mean free path phonons. The estimated average $\lambda$-value for these phonons was around 4.5 nm, much higher than thermal phonons at room-temperature that have $\lambda \sim 0.2$nm. We thus believe that the main difference between both results arises from a different microstructure of the used $SiN_x$ in each case. The thermal conductivity value of our $SiN_x$, slightly below Sultan et al.[22,25] values combined with the temperature dependence shown in Fig. 4, clearly indicates that our nitride is fully amorphous while the one used in Sultan et al work was probably nanocrystalline. Therefore, a lower drop of the thermal conductance in fully amorphous nitride membrane is consistent with a scenario where the contribution of long-$\lambda$ phonons is reduced with respect to previous nanocrystalline nitride membranes. Even though, the conductance drop due to enhanced surface scattering requires that phonons slightly larger than the average interatomic distance at room temperature have to be involved in heat transport along the nitride membrane. We can thus tentatively attribute the initial sharp reduction of the thermal conductance to the formation of TPD isolated clusters on top of the nitride membrane surface, modifying the interfacial phonon scattering and thus leading to a decrease of the thermal conductance.

Although we currently lack a complete understanding of the microscopic processes occurring at the interfaces, we believe that the growth of new material on top of the membrane changes the specularity of the surface, resulting on an effective increase in the phonon scattering rate. We foresee that the future use of crystalline membranes, such as single crystalline Si, will provide a convenient platform to investigate nucleation and island growth dynamics during the early stages of film growth with even higher conductance sensitivity.

In Region II the thermal conductance still decreases but with a lower slope. We believe that in this region, clusters enlarge and start to coalesce, providing additional paths for heat transfer which partially compensate the interface scattering until a minimum is reached at the end of the present region. As island coalescence continues, percolation builds up new channels across the layer structure (Region III), providing additional heat flow paths that start to exceed the contribution of the interface scattering. The coincidence of the minimum in thermal conductance with a percolation phenomenon threshold was demonstrated in a separate experiment where electrical conductance was measured as a function of thickness (inset in Fig. 5b), showing a sharp variation of the slope at 2.6 nm, due to electrical continuity through the TPD film.

As percolation persists, the thermal conductance increases reaching a linear regime that we identify with the formation of a continuous layer with complete coverage, marking the onset of Region IV. Thus, Region IV corresponds to the vertical growth of the continuous film. In this regime the increase in thermal conductance is linear and proportional to the thickness of the growing layer. Compared to the end of region III, there is a small reduction in the slope of the conductance vs thickness since the islands are no longer forming new conductive channels.

Differences in growth dynamics with the deposition temperature were also studied, as shown in Fig. 5c, where TPD film was deposited at 304K. Although the four regions appear in both cases, remarkable differences in region limits appears compared with sample deposited at 267 K (Fig. 5b). The TPD sample grown at $T_{dep}$=267 K shows values of the percolation threshold (black downward arrows) and film continuity at 2.5 nm and 14 nm, respectively. However, sample grown at 304 K showed higher percolation threshold, 6 nm (probably due to higher molecular mobility taking place at a higher deposition temperature), and the thickness value for film continuity is not clearly resolved from the data since the reduction in the slope after Region III is not observed. Recent work by Fakhraai and coworkers[26] in TPD films grown at 315K have shown that film continuity was reached for film thickness above 20 nm, which is consistent with our results.

Although similar behaviour of $K_{2\omega}$ was observed for both samples, final values of in-plane conductivity determined form the slope in Regions IV of Fig. 5a and c, differ considerably: $k$=0.145 Wm$^{-1}$K$^{-1}$ for films deposited at $T_{dep}$=267 K and $k$= 0.132 Wm$^{-1}$K$^{-1}$ for those deposited at $T_{dep}$=304 K. While the difference only amounts to 10%, it is substantially larger than the measurement uncertainty. Both films were expected to be amorphous at those deposition temperatures, as was confirmed by clear glass transition signatures shown in the calorimetric traces of Fig. S6 (SI). We thus believe that the variation in thermal properties has to be related to the diverse characteristics of vapour-deposited glasses, in particular density and molecular orientation, which strongly depend on the deposition temperature. It has already been demonstrated that vapour-deposited thin-film organic glasses grown at deposition temperatures slightly below their glass transition develop enhanced kinetic and thermodynamic stability with a maximum at $T_{dep}$ in the vicinity of $0.85T_g$[27,28]. Glasses grown in these conditions are coined ultrastable glasses. The glass transition temperature of a conventional TPD glass (this is, a glass cooled from the liquid at 10 K/min) is 333 K. Glasses grown in the region 0.80-0.90$T_g$, which is the case for film deposited at 267 K, are stable glasses, as evidenced by the higher onset of their glass transition temperature upon heating(Fig. S6 in SI). However, films grown above 0.90$T_g$, which is the case for film grown at 304K, are less stable. Stability can be directly translated to density, meaning that the sample grown at 267 K (0.80$T_g$) has a slightly higher density that the one vapor-deposited at 304 K (0.91$T_g$). According to Dalal et al.[29], the difference in density between the 2 samples should be around 0.3%. Besides, stable glasses embedded with higher density also exhibit higher values of the sound velocity up to 10%[30–32], therefore it is reasonable to expect that more stable glasses will also exhibit an enhancement of the thermal conductivity.

In the same way, vapour-deposited stable glasses show anisotropic molecular packing with a molecular orientation that depends on the deposition temperature. TPD films grown at $0.80T_g$ have molecules partially aligned parallel to the substrate while those grown at $0.91 T_g$ are mostly randomly oriented[29]. Molecular anisotropy can also play a role in heat transport since it could be slightly favoured in the direction of molecular alignment. More studies of the thermal conductivity variation as a function of deposition temperature are under way to disentangle the effects of density and molecular orientation.

Metallic thin-film layers

To complete the proof of concept of the sensor, we also analyse growth kinetics of a metallic indium layer. Fig.6 shows the real-time in-situ thermal conductance measurement during growth, also for two deposition temperatures, 315 K and 260 K. Conductance values (Fig. 6a,b) follow a similar pattern to the one observed in TPD films but with much larger thickness scales. The conductance regions can be identified and discussed, as in the TPD case, in view of microstructure evolution with thickness (see Fig. S4 for a complete set of microscopy images). As shown in the inset of Fig. 6a, the fast decrease of conductance at the very early stages of deposition is also present. In this thickness range (up to 0.65 nm), tiny isolated clusters are expected to develop on the surface of the $SiN_x$ membrane. The observed conductance drop of 1.5% is again an indication that phonons with mean-free-paths slightly larger than those typically accounted for in disordered solids are being scattered by the In/$SiN_x$ interface. Nevertheless, the minimum conductance found in this case cannot be correlated with a percolation threshold, as we considered in TPD, since individual In islands are much more conductive than the TPD ones and they can contribute significantly to the thermal conduction, even though they are still physically disconnected at this stage.

Interestingly, neither percolation threshold nor complete coverage of In grown on $SiN_x$ are reached for metallic layer as thick as 120 nm (nominal thickness), as clearly evidenced in the SEM images. As shown in Fig.6b, where growth characterization is performed up to 450 nm, percolation starts to play a significant role at thicknesses around 200 nm, where the conductance sharply increases due to continuous channels formation. Subsequently, the thermal conductance increase slows down becoming linear with thickness. The corresponding fitting line (red line in Fig. 6b) is proportional to the thickness, meaning that the In layer is already continuous and the thermal conductivity can be evaluated from the slope of the conductance versus thickness. The value of $k$ for the continuous In layer deposited at 260K is 47 $Wm^{-1}K^{-1}$, substantially lower than the tabulated value for In, $k$=83 Wm$^{-1}$K$^{-1}$, probably due to smaller grain size and enhanced boundary scattering. From the images taken by SEM of the In layer at different thicknesses (insets of Fig.6a and SI Fig.S4b), we approximate the coverage ratio and the mean island area, as specified in SI.

As indicated in Fig. 6a,b we can differentiate four growth regimes: I: nucleation and initial small island growth. II: growth of islands, divided in IIa (with small isolated circular islands) and IIb (with larger islands irregularly shaped forming a bimodal island size distribution). III: island percolation forming continuous channels and IV: vertical growth of a continuous film.

Since the conductance evolution is slightly different (region II is divided in two stages) than the one previously observed for TPD layers, we conduct finite element modelling using the structural information provided by the electron microscopy images. We use a simplified representation of our sample by building an array of 9x9 In square islands on a 180 nm thick SiN$_x$ film. Changing the nominal thickness of the layer implies modulating the size and separation of the islands to match the island size and coverage ratio observed by SEM (Table S2 in SI). The thermal conductance was monitored by imposing a heat flow and measuring the temperature difference arising on the simulated structure. The results of the simulation are shown in Fig.6c. The simulation closely predicts the increase of the slope of the curve $K_{2\omega}(t_{smp})$ deposited at 315 K in a thickness range around 30-50 nm. In this thickness range and up to the percolation threshold above 200 nm In islands are still isolated from one another and the increased conductance is due to the formation of larger islands as In evaporation proceeds. At around 200 nm the sudden increase in the slope is related to island percolation (Region III starts). At the end of the percolation regime a continuous film forms and the conductance is heavily dominated by the In film (Region IV). Then, the conductance increases linearly with a slope given by the thermal conductivity of the film.

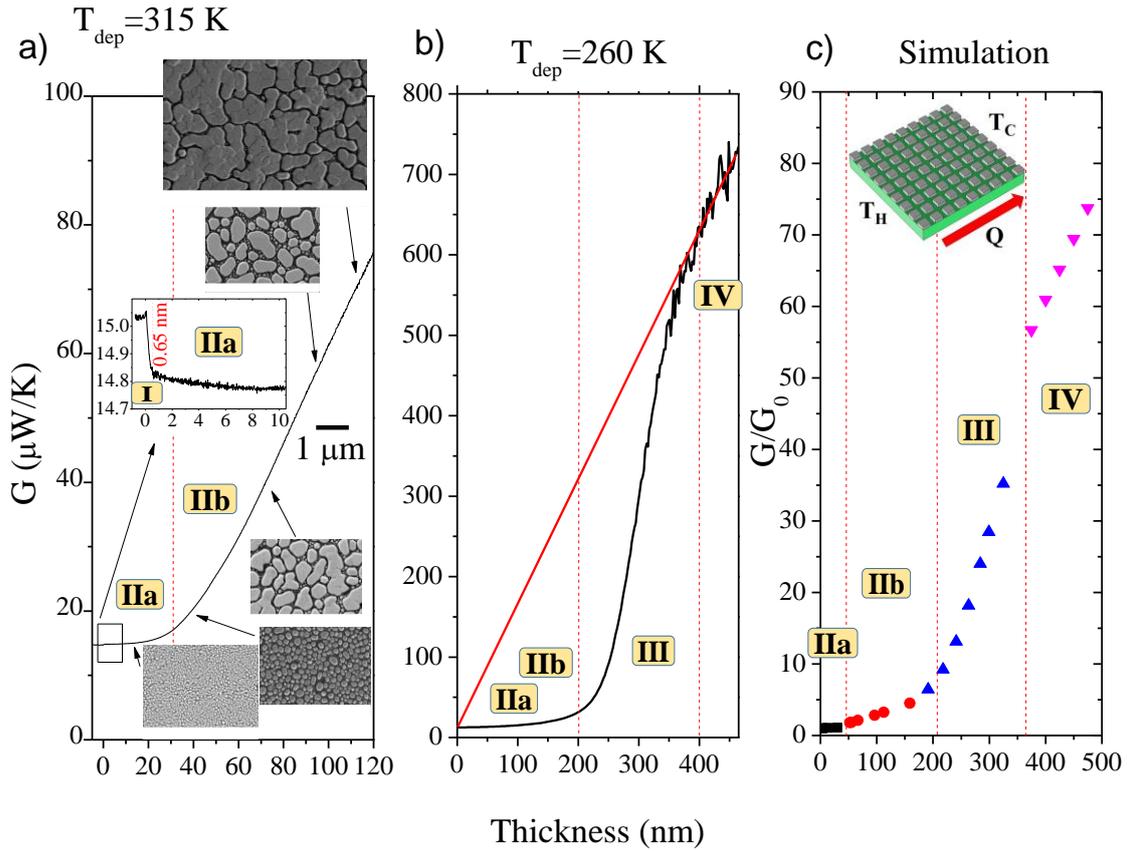

**Fig.6** a) Thermal conductance vs thickness during deposition of In films: a) $T_{dep}$=315 K. The electron microscopy images of films of varying nominal thickness show the microstructural evolution of the In layer. SEM images were acquired with a magnification of 100k, except the thickest one recorded at 50k. The 1 μm scale is provided inside the graph. The inset of a) highlights the conductance drop during the early growth stages. b) $T_{dep}$=260 K. Thickness range is extended to attain complete percolation and total coverage. c) FEM simulation: normalized conductance vs. In nominal thickness for representative structures with isolated islands of different sizes and percolated islands, finally forming a continuous layer. The different growth regimes are shown in Roman letters. Inset of c): Image of the 3D model, consisting on an array of 9x9 square In islands (grey) on a SiN$_x$ thin membrane(green).

## Conclusions

We have developed a universal sensor with extreme accuracy and methodology able to perform real-time thermal conductance measurements during film growth. The technique is adapted for any kind of material and thickness. Large sensitivity allows disentangling the evolution of film microstructure with thicknesses. By analysing changes in the apparent thermal conductance versus thickness, differentiated growth stages could be identified giving rise to a complete comprehension of the growth process for several materials. At early stages, the fast drop of the thermal conductance was related to nucleation and isolated clusters formation. This step is followed by a regime where clusters grow to form islands through coalescence and absorption of atoms/molecules from the gas phase, which dominates the variation of the heat conductance with thickness. In an intermediated stage the percolation threshold was revealed by a conductance rise while final mode, where the thermal conductance changes linearly with thickness, corresponded to the formation and growth of a continuous film. In this regime, the thermal conductivity of the film can be directly derived from the slope of the conductance versus thickness plot.

The methodology ad-hoc, presented here, is easily extensible to devices with other substrate materials compatibles with epitaxial growth. Using single crystalline membranes, the conductance reduction within the first stages of growth will be enhanced increasing the sensitivity. The extreme sensitivity will pave the way to apply the technique to interesting phenomena such as phase changes during growth, size effects or molecular orientation and density in organic glasses films, among others.

.

## Acknowledgements

We acknowledge financial support from the Ministerio de Economía y Competitividad through Grants CSD2010-00044 (ConsoliderNANOTHERM), FIS2013-50304-EXP and MAT2016-79579-R. P. Ferrando-Villalba and J. Ràfols-Ribé were in receipt of an FPU grant from the Spanish Ministry of Education, Culture and Sports. Ll. Abad was supported by the "Ramón y Cajal" program from Spanish Government. The authors gratefully acknowledge the ICTS-IMB-CNM clean room for the chips microfabrication.

# Additional information

## Author Contributions Statement

## Competing financial interests.

The author(s) declare no competing financial interests.